\documentclass[hyper]{JHEP} 

\usepackage{epsfig}

\newcommand\fverb{\setbox\pippobox=\hbox\bgroup\verb}

\newcommand\fverbdo{\egroup\medskip\noindent%

            \fbox{\unhbox\pippobox}\ }

\newcommand\fverbit{\egroup\item[\fbox{\unhbox\pippobox}]}

\newbox\pippobox

\title{Note About Canonical Description of T-duality
Along Light-Like Isometry}
\author{J. Kluso\v{n}\\
Department of
Theoretical Physics and Astrophysics\\
Faculty of Science, Masaryk University\\
Kotl\'{a}\v{r}sk\'{a} 2, 611 37, Brno\\
Czech Republic\\
E-mail: \email{klu@physics.muni.cz}} \preprint{}

 \abstract{In this short note we analyze canonical description of
 T-duality along  light-like isometry. We show
 that T-duality of
relativistic string theory on this background
leads to non-relativistic
string theory action on T-dual background.}

\def\tp{\tilde{p}}

\def\hG{\hat{G}}

\def\hB{\hat{B}}
\def\bA{\mathbf{A}}

\def\bV{\mathbf{V}}

\def\ty{\tilde{y}}

\def\tlambda{\tilde{\lambda}}
\def\bB{\mathbf{B}}

\def\tx{\tilde{x}}

\def\be{\begin{equation}}

\def\ee{\end{equation}}

\def\bea{\begin{eqnarray}}

\def\eea{\end{eqnarray}}

\def\tmH{\tilde{\mH}}

\def\bY{\mathbf{Y}}

\def\mH{\mathcal{H}}

\def\tG{\tilde{G}}

\def\bM{\mathbf{M}}

\newcommand{\mG}{\mathcal{G}}

\def \bA{\mathbf{A}}

\newcommand{\mL}{\mathcal{L}}

\def\pb #1{\left\{#1\right\}}

\begin{document}
\section{Introduction and Summary}
It is well known that two string theories,
one defined on the background with  compact dimension of radius $R$, and the second one defined on background with compact dimension of  radius $R'=\frac{\alpha'}{R}$, are equivalent.
This duality has its origin in the extended nature of the string when we can exchange
momentum with winding numbers, respectively. The most powerful description of such
a duality is given in terms of Buscher's rules
\cite{Buscher:1987sk,Buscher:1987qj} of the transformations of the background fields under
T-duality. More explicitly, we start with string sigma model on the background where the
background metric possesses one isometry. Then we gauge this isometry so that this is now local symmetry on the string world-sheet when we introduce corresponding
covariant derivative and two dimensional gauge field. Since this gauge field has to  be non-dynamical in order not to change physical content of the theory we  add to the action term that ensures that
the field strength of this gauge field is zero on shell. As the next step we fix the gauge when we take the world-sheet mode that parameterizes direction with gauged isometry to be zero. Then we can solve the flatness of the gauge field by introducing  a new scalar mode
that parameterizes the string propagating along dual coordinate where now the background
fields are related to the original ones through Buscher's rules.

There  is an alternative approach to this standard treatment of T-duality which  is description of T-duality with the help of canonical transformations \cite{Alvarez:1994wj,Alvarez:1994dn}. This procedure is based on the
Hamiltonian form of string in the background that possesses an isometry.  Then we perform
specific canonical transformations of the world-sheet mode and its conjugate momenta that
correspond to coordinate along this isometry. As a result we obtain
a new Hamiltonian for dual theory. Finally we perform inverse Legendre transformations to T-dual Lagrangian with the background fields again determined by Buscher's rules.

While this procedure is well established for space-like isometry  the case of light-like isometry is much less known and analyzed. The aim of this short note is to
focus on this problem, inspired by recent analysis of
 T-duality in the context of non-relativistic string theories
\cite{Bergshoeff:2018yvt,Kluson:2018vfd}, where non-relativistic strings were firstly introduced in seminal papers \cite{Gomis:2000bd,Danielsson:2000gi}
\footnote{For related works, see for example
  \cite{Gomis:2005pg,Gomis:2019zyu,Andringa:2012uz,Harmark:2017rpg,Kluson:2018egd,Kluson:2018grx,Harmark:2018cdl,Kluson:2019ifd}.}. In fact, as we will show explicitly in the
next section, the canonical analysis of T-duality for string with light-like isometry
leads to T-dual action where the kinetic term for dual coordinate is missing. Then in order to solve this problem we introduce two auxiliary fields $\lambda^+$ and $\lambda^-$ in such a way that the original Lagrangian is quadratic in coordinate that we dualize. However we still have to ensure that solving equations of motion for $\lambda^+,\lambda^-$ we return to the original Lagrangian density.
When we have an extended Lagrangian density we can find its Hamiltonian form and then we perform canonical transformation corresponding T-duality along original light-like coordinate. From this T-dual Hamiltonian we derive corresponding Lagrangian density and we find that the background fields transform according to Buscher's rules. The most remarkable fact considering T-dual Lagrangian density is that it is linear in auxiliary fields $\lambda^+,\lambda^-$ that is characteristic property of non-relativistic string theory action
\cite{Gomis:2000bd,Bergshoeff:2018yvt,Gomis:2019zyu}. To see this in more details we solve the equations of motion for $\lambda^+,\lambda^-$  that imply that Lagrangian density has the form
of non-relativistic string with specific induced world-sheet metric. Clearly for the form of  background that was studied in \cite{Bergshoeff:2018yvt,Kluson:2018vfd} the non-relativistic string corresponds to string in stringy Newton-Cartan background.

We mean that this result is very interesting since it shows that string theory on the background with light-like isometry is T-dual to specific form of non-relativistic string. This could be very useful for the quantum description of string theory on the background with light-like isometry, following recent interesting paper  \cite{Gomis:2000bd}. It would be also very interesting to extend the canonical analysis of T-duality transformation presented in this paper to the case of
Green-Schwarz superstring. We hope to return to this problem in future.

This paper is organized as follows. In the next section (\ref{second}) we review canonical treatment of T-duality and we show difficulty with its application on the case of string in the background with light-like isometry. Then in section (\ref{third})  we perform T-duality transformation in case of the extended action and we show that it leads to non-relativistic string in T-dual background.

\section{T-duality in Canonical Formalism}\label{second}
In this section we review basic facts about T-duality as a canonical transformations, following   \cite{Alvarez:1994wj,Alvarez:1994dn}. Let us be more concrete with the definition of the canonical formalism in case of the bosonic string. We have two sets of canonical variables: $p_\mu, x^\mu$ where $\mu,\nu=0,\dots,25$ and their duals: $\tp_\mu,\tx^\mu$ where we demand that both canonical variables give equivalent equations of motions. In other words we demand that the string action can be written in two equivalent ways
\begin{equation}
S=\int d\tau d\sigma (p_\mu\partial_\tau x^\mu-\mH)=
\int d\tau d\sigma (\tp_\mu\partial_\tau \tx^\mu -\tmH)+\int d\tau\frac{dG}{d\tau} \ ,
\end{equation}
where $G$ is generating function of canonical transformations and where we label world-sheet of the string with two coordinates $\tau$ and $\sigma$.
 It is well known that there are four possible forms of the generating function that differ in dependence on two sets of canonical variables. In our case we presume that they depend on original and dual variables, in other words $G=G(x,\tx)$. We presume that $G=\int d\sigma\mG$ and hence we can write
\begin{equation}
p_\mu \partial_\tau x^\mu-\mH(x^\mu,p_\mu)=
\tp_\mu \partial_\tau \tx^\mu-\tmH(\tx^\mu,\tp_\mu)+
\frac{\partial \mG}{\partial \tau}+\frac{\delta \mG}{\delta x^\mu}\partial_\tau x^\mu+
\frac{\delta \mG}{\delta \tx^\mu}\partial_\tau \tx^\mu
\end{equation}
so that comparing we obtain relation
\begin{eqnarray}
p_\mu=\frac{\delta \mG}{\delta x^\mu} \ , \quad \tp_\mu=-\frac{\delta \mG}{\delta \tx^\mu} \ , \quad
\tmH(\tx^\mu,\tp_\mu)=\mH(x^\mu,p_\mu)+\frac{\partial \mG}{\partial\tau} \ .  \nonumber \\
\end{eqnarray}
This is general form of canonical transformation for bosonic string. In case of T-duality transformation we presume that $\mG$ does not explicitly depend on $\tau$. Further, we perform
canonical transformation with respect to one coordinate, say $x^{25}\equiv y$ that labels isometry direction. In this case the generating function has the form
\cite{Alvarez:1994wj,Alvarez:1994dn}
\begin{equation}
G(y,\ty)=\frac{T}{2}\int d\sigma (\partial_\sigma y\ty-y\partial_\sigma \ty) \ ,
\end{equation}
where $T=\frac{1}{2\pi\alpha'}$ is string tension.
Let us denote momentum conjugate to $\ty$ as $p_{\ty}$. Then from
the definition of the canonical transformations we derive following
relation between $\ty$ and $p_{\ty}$ in the form
\begin{eqnarray}
& &p_{\ty}=-\frac{\delta G}{\delta \ty}=-T\partial_\sigma y \ , \nonumber \\
& &p_y=\frac{\delta G}{\delta y}=-T\partial_\sigma \ty \ . \nonumber \\
\end{eqnarray}
With the help of these relations we obtain dual Hamiltonian when we replace $
\partial_\sigma y$ with $-\frac{1}{T}p_{\ty}$ and $p_y$ with $-T\partial_\sigma y$.
Let us now be more explicit and consider
  Polyakov string action in background with light-like isometry along $y$ direction which means
  that all background fields do not depend on $y$ explicitly and also that the
  metric component $G_{yy}$ is absent. Then the Lagrangian has the form
\begin{equation}
\mL=-\frac{T}{2}\sqrt{-\gamma}[\gamma^{\alpha\beta}\partial_\alpha x^i
\partial_\beta x^j G_{ij}+2\gamma^{\alpha\beta}\partial_\alpha x^i G_{iy}\partial_\beta y]-\frac{T}{2}\epsilon^{\alpha\beta}B_{\mu\nu}\partial_\alpha x^\mu
\partial_\beta x^\nu \ ,
\end{equation}
where $\epsilon^{\alpha\beta}=-\epsilon^{\beta\alpha} \ , \epsilon^{\tau \sigma}=1$ and where $i,j=0,\dots,24$.  Let us
introduce following $1+1$ parameterization of the world-sheet metric $\gamma_{\alpha\beta}$ as
\begin{equation}
\gamma_{\alpha\beta}=\left(\begin{array}{cc}
-N^2+N^\sigma\omega N^\sigma & N^\sigma\omega \\
N^\sigma\omega & \omega \\
\end{array}\right) \ ,
\gamma^{\alpha\beta}=
\left(\begin{array}{cc}
-\frac{1}{N^2} & \frac{N^\sigma}{N^2} \\
\frac{N^\sigma}{N^2} & \frac{1}{\omega}-
\frac{N^\sigma N^\sigma}{N^2} \\ \end{array}\right)
\end{equation}
so that the Lagrangian has the form
\begin{eqnarray}\label{mLoriginal}
& &\mL=-\frac{T}{2}N\sqrt{\omega}
[-\nabla_n x^i G_{ij}\nabla_n x^j+\frac{1}{\omega}
\partial_\sigma x^i\partial_\sigma x^jG_{ij}-2\nabla_n x^i
G_{iy}\nabla_n y+\frac{2}{\omega}\partial_\sigma \tx^iG_{iy}\partial_\sigma y]-\nonumber \\ & &-TB_{\mu\nu}\partial_\tau x^\mu \partial_\sigma x^\nu \ , \quad \nabla_n\equiv
\frac{1}{N}(\partial_\tau-N^\sigma \partial_\sigma) \ .   \nonumber \\
\end{eqnarray}
From (\ref{mLoriginal})  we obtain conjugate momenta
\begin{eqnarray}
& &p_i=T\sqrt{\omega}G_{ij}\nabla_nx^j+T\sqrt{\omega}G_{iy}
\nabla_n y-TB_{i\mu}\partial_\sigma x^\mu \ , \quad
p_{y}=T\sqrt{\omega}G_{y i}\nabla_n x^i-TB_{y\mu}\partial_\sigma x^\mu  \ . \nonumber \\
& & p_N\approx 0 \ , \quad p_\omega\approx 0 \ , \quad p_{N^\sigma}\approx 0 \ , \nonumber \\
\end{eqnarray}
where $p_N,p_\omega,p_{N^\sigma}$ are momenta conjugate to $N,\omega$ and $N^\sigma$, respectively.
Then performing  standard analysis we obtain the  Hamiltonian density in the form
\begin{eqnarray}\label{mHnaive}
& &\mH=\frac{N}{2T\sqrt{\omega}}\left(\pi_i
G^{ij}\pi_j+
\pi_{y}G^{yy}\pi_{y}+
2\pi_{y}G^{y i}\pi_i+\right.\nonumber \\
& &\left.+T^2G_{ij}\partial_\sigma x^i
\partial_\sigma x^j+2T^2\partial_\sigma x^i
G_{iy}\partial_\sigma y\right)
+N^\sigma (p_i \partial_\sigma x^i+p_{y}\partial_\sigma y) \ ,
\nonumber \\
\end{eqnarray}
where  $\pi_\mu=p_\mu+TB_{\mu\nu}\partial_\sigma x^\nu$ and where the inverse metric $G^{\mu\nu}$ has  following form
\begin{equation}
G^{yy}=-\frac{1}{G_{y i}\tG^{ij}G _{jy}} \ , \quad
G^{y j}=\frac{G_{y k}\tG^{kj}}{G_{y i}G^{ij}
G_{jy}} \ , \quad
G^{ij}=\tG^{ij}-\frac{G_{y k}\tG^{ki}G_{y l}\tG^{lj}}{G_{y i}\tG^{ij}
    G_{jy}} \ ,
\end{equation}
where $\tG^{ij}$ is metric inverse to $G_{ij}$ so that
$\tG^{ik}G_{kl}=\delta^i_l$.  From (\ref{mHnaive}) we also see that it is convenient
to introduce $N^\tau=\frac{N}{2T\sqrt{\omega}}$ so that $\omega$ disappears from the Hamiltonian
as expected.

Then following previous general discussion we perform T-duality in (\ref{mHnaive}) when
\begin{equation}
p_{\ty}=-T\partial_\sigma y \ ,  \quad p_y=-T\partial_\sigma \ty
\end{equation}
so that
\begin{eqnarray}
\pi_y=-T\partial_\sigma \ty+TB_{y i}\partial_\sigma x^i=-T\bV \ , \quad
\pi_i=p_i+T B_{ij}\partial_\sigma x^j+TB_{iy}\partial_\sigma y=
k_i-B_{iy} p_{\ty} \ . \nonumber \\
\end{eqnarray}
Inserting these results into (\ref{mHnaive}) we obtain
 T-dual Hamiltonian constraint in the form
\begin{eqnarray}\label{mHTnaive}
& &\mH^T_\tau=T^2 \bV G^{yy}\bV-2T(k_i-B_{iy}p_{\ty})G^{iy}\bV+
(k_i-B_{iy}p_{\ty})G^{ij}(k_j-B_{jy}p_{\ty})+\nonumber \\
& &+T^2G_{ij}\partial_\sigma x^i\partial_\sigma  x^j-2T G_{iy}\partial_\sigma x^i p_{\ty} \ . 
\nonumber \\
\end{eqnarray}
We see that for zero NSNS two form the Hamiltonian constraint is
linear in $p_{\ty}$ which suggests possible difficulties with this
theory. This can be also seen even in case of non-zero $B_{\mu\nu}$
when we determine corresponding Lagrangian.
 In fact, using
(\ref{mHTnaive}) in the T-dual Hamiltonian
 $H^T=\int d\sigma (N^\tau \mH^T_\tau
+N^\sigma \mH^T_\sigma)$ where $\mH_\sigma=p_i \partial_\sigma x^i +
p_{\ty}
\partial_\sigma \ty$ we obtain time derivatives of $x^i$ and $\ty$ as
\begin{eqnarray}
& &\partial_\tau x^i=\pb{x^i,H^T}=
2N^\tau G^{ij}(k_j-B_{jy}p_{\ty})-2N^\tau T G^{iy}\bV+N^\sigma \partial_\sigma x^i \ , \nonumber \\
& &\partial_\tau \ty=\pb{\ty,H^T}=
2N^\tau T B_{i y}G^{i y}\bV-2N^\tau B_{i y}G^{ij}(k_j-B_{j y}p_{\ty})
-2N^\tau T G_{i y}\partial_\sigma x^i+N^\sigma \partial_\sigma \ty
\nonumber \\
\end{eqnarray}
and hence we  find
\begin{eqnarray}
Y=-B_{i\ty}X^i-2N^\tau T G_{i\ty}\partial_\sigma x^i \ , \quad
Y=\partial_\tau \ty-N^\sigma\partial_\sigma \ty \ , \quad
X^i=\partial_\tau x^i-N^\sigma \partial_\sigma x^i \
\nonumber \\
\end{eqnarray}
which means that there is no relation between $\partial_\tau \ty$ and conjugate momenta.
In fact, if we proceed further we obtain naive T-dual Lagrangian density in the form
%
%
\begin{eqnarray}
& &\mL^T_{naive}=\frac{1}{4N^\tau}X^iG_{ij}X^j+T
G^{i y}\bV G_{ij}X^j+\nonumber \\
& &+N^\tau T^2 G^{i y}\bV G_{ij}
G^{j y}\bV-N^\tau T^2 \bV G^{yy}\bV-N^\tau
T^2 G_{ij}\partial_\sigma x^i\partial_\sigma x^j-T
B_{ij}\partial_\tau x^i\partial_\sigma x^j\nonumber \\
\end{eqnarray}
and we see that the kinetic term for T-dual variable $\ty$ is absent which is
not satisfactory result. In the next section we will try to resolve this puzzle by introducing
two auxiliary fields.
\section{Extended Lagrangian and T-duality}\label{third}
In order to resolve this puzzle let us introduce an equivalent form of the
Lagrangian density to the original one
(\ref{mLoriginal}). To do this we introduce two auxiliary fields
$\lambda^+$ and $\lambda^-$  and consider following Lagrangian density
\begin{eqnarray}\label{mLextended}
& &\mL=-\frac{T}{2}N\sqrt{\omega}
[-\nabla_n x^i \hG_{ij}\nabla_n x^j+\frac{1}{\omega}
\partial_\sigma x^i\partial_\sigma x^j\hG_{ij}-2\nabla_n x^i
\hG_{iy}\nabla_n y+\frac{2}{\omega}\partial_\sigma x^i\hG_{iy}\partial_\sigma y-\nonumber \\
& &-\nabla_n y \hG_{yy}\nabla_n y+\frac{1}{\omega}
\partial_\sigma y \hG_{yy}\partial_\sigma y+
\lambda^+ \bA+\lambda^-\bB+\lambda^+\lambda^-]-T\hat{B}_{\mu\nu}
\nabla_n x^\mu \partial_\sigma x^\nu \ ,
\nonumber \\
\end{eqnarray}
where now we have to choose $\bA$ and $\bB$ in such a way to ensure
that $\hG_{\ty\ty}=0$ after solving equations of motion for $\lambda^+$ and
$\lambda^-$. Explicitly, from (\ref{mLextended}) we find  their equations of motion in the form
\begin{equation}
\lambda^-=-\bA \ , \quad \lambda^+=-\bB \ .
\end{equation}
Inserting these results into (\ref{mLextended}) we obtain that they give following
contribution  to the Lagrangian density
\begin{equation}
\lambda^+\bA+\lambda^-\bB+\lambda^+\lambda^-=-\bA\bB \ .
\end{equation}
As the next step we presume that  $\bA$ and $\bB$ have the form
\begin{eqnarray}
\bA=\nabla_n x^i \bA_i+\nabla_n y \bY^+-\frac{1}{\sqrt{\omega}}
[\partial_\sigma x^i\bA_i+\partial_\sigma y \bY^+] \ ,
\nonumber \\
\bB=\nabla_n x^i \bB_i+\nabla_n y \bY^-+\frac{1}{\sqrt{\omega}}
[\partial_\sigma x^i\bB_i+\partial_\sigma y \bY^-] \ ,
\nonumber \\
\end{eqnarray}
so that
\begin{eqnarray}\label{AB}
& &\bA \bB=
\nabla_n x^i\nabla_n x^j\bA_i\bB_j+\nabla_n x^i\nabla_n y
(\bA_i\bY^-+\bY^+\bB_i)+\nonumber \\
& &+\frac{2}{
\sqrt{\omega}}\nabla_n x^i\partial_\sigma x^j(\bA_i\bB_j-\bA_i\bB_j)
+\frac{1}{\sqrt{\omega}}\nabla_n x^i \partial_\sigma y(\bY^-\bA_i-
\bY^+\bB_i)+
\nonumber \\
& &+\nabla_n y\nabla_n y \bY^+\bY^-+\frac{1}{\sqrt{\omega}}\nabla_n y\partial_\sigma x^i
(\bY^+\bB_i-\bY^-\bA_i)
-\nonumber \\
& &-\frac{1}{\omega}(\partial_\sigma x^i\partial_\sigma x^j \bA_i\bB_j+
\partial_\sigma x^i\partial_\sigma y(\bA_i\bY^-+\bB_i\bY^+)+
\partial_\sigma y\partial_\sigma y \bY^+\bY^-) \ .
\nonumber \\
\end{eqnarray}
For simplicity we begin with our analysis with the  minimal case when $\bA_i=\bB_i=0$.
\subsection{Minimal Case}
First of all we should demand that  when we solve the equations of motion for
$\lambda^+,\lambda^-$ the Lagrangian (\ref{mLextended}) with additional contribution given in
(\ref{AB})
reduces to the original one
with $G_{yy}=0$. This condition implies
\begin{equation}
\hG_{yy}+\bY^+\bY^-=0 \
\end{equation}
that can be solved as $\bY^+=\sqrt{\hG_{yy}} \ , \bY^-=-\sqrt{\hG_{yy}}$. In this case (\ref{AB})
is equal to
\begin{eqnarray}
\bA\bB=[\nabla_n y \nabla_n y-\frac{1}{\omega}\partial_\sigma y
\partial_\sigma y]\bY^+\bY^- \ .
\end{eqnarray}
Inserting this result into (\ref{mLextended}) and
comparing  with (\ref{mLoriginal})
we obtain following relation between original metric and NSNS two form fields and
hatted ones:
\begin{equation}
G_{ij}=\hG_{ij} \ , \quad G_{iy}=\hG_{iy} \ , \quad B_{\mu\nu}=
\hB_{\mu\nu} \ .
\end{equation}
Now we  return to  (\ref{mLextended}) that for $\bA_i=\bB_i=0$ has the form
\begin{eqnarray}
& &\mL=-\frac{T}{2}N\sqrt{\omega}
[-\nabla_n x^i \hG_{ij}\nabla_n x^j
-\nabla_n y \hG_{yy}\nabla_n y
-2\nabla_n x^i
\hG_{iy}\nabla_n y+\nonumber \\
&+&
\frac{1}{\omega}
\partial_\sigma x^i\partial_\sigma x^j\hG_{ij}
+
\frac{2}{\omega}\partial_\sigma x^i\hG_{iy}\partial_\sigma y+\frac{1}{\omega}
\partial_\sigma y \hG_{yy}\partial_\sigma y+\nonumber \\
& &+\lambda^+(\nabla_n y-\frac{1}{\sqrt{\omega}}\partial_\sigma y)\bY^++\lambda^-
(\nabla_n y+\frac{1}{\sqrt{\omega}}\partial_\sigma y)\bY^-
+\lambda^+\lambda^-]-T\hat{B}_{\mu\nu}
\nabla_n x^\mu \partial_\sigma x^\nu \ .
\nonumber \\
\end{eqnarray}
As the next step we proceed to the canonical formalism. From Lagrangian density given  above
we obtain
\begin{eqnarray}
& &p_i=T\sqrt{\omega}[\hG_{ij}\nabla_n x^j+
\hG_{iy}\nabla_n y]-T\hat{B}_{i\nu}\partial_\sigma x^\nu \ , \nonumber \\
& & p_y=T\sqrt{\omega}
[\hG_{yy}\nabla_n y+\hG_{yi}\nabla_n x^i -\frac{1}{2}\lambda^+\bY^+
-\frac{1}{2}\lambda^-\bY^-]
-T \hat{B}_{yi}\partial_\sigma x^i \  \nonumber \\
& & p_N\approx 0 \ ,\quad p_{\omega}\approx 0 \ , \quad p_{N^\sigma}\approx 0 \
\nonumber \\
\end{eqnarray}
and hence Hamiltonian density has the form
\begin{eqnarray}
& &\mH=\frac{N}{2\sqrt{\omega}T}
\left(\pi_i \hG^{ij}\pi_j+2\pi_i \hG^{iy}
(\pi_y+\frac{T}{2}\sqrt{\omega}\lambda^+\bY^++
\frac{T}{2}\sqrt{\omega}\lambda^-\bY^-)+\right.\nonumber \\
& &\left.+(\pi_y+\frac{T}{2}\sqrt{\omega}\lambda^+\bY^++
\frac{T}{2}\sqrt{\omega}\lambda^-\bY^-)\hG^{yy}
(\pi_y+\frac{T}{2}\sqrt{\omega}\lambda^+\bY^++
\frac{T}{2}\sqrt{\omega}\lambda^-\bY^-)\right)+\nonumber \\
& &+\frac{TN}{2\sqrt{\omega}}\partial_\sigma x^i\hG_{ij}\partial_\sigma x^j
+\frac{TN}{\sqrt{\omega}}\partial_\sigma x^i \hG_{ij}\partial_\sigma y+
\frac{TN}{2\sqrt{\omega}}\partial_\sigma y \hG_{yy}\partial_\sigma y-\nonumber \\
& &-\frac{TN}{2}\lambda^+\partial_\sigma y \bY^++\frac{TN}{2}
\lambda^-\partial_\sigma y \bY^-+\frac{T}{2}N\sqrt{\omega}\lambda^+\lambda^-+
N^\sigma \mH_\sigma \ .
\nonumber \\
\end{eqnarray}
Performing  rescaling
\begin{equation}
\sqrt{\omega }\lambda^+=\tlambda^+ \ , \quad
\sqrt{\omega}\lambda^-=\tlambda^-
\end{equation}
and introducing $N^\tau=\frac{N}{2\sqrt{\omega}T}$ we can rewrite the Hamiltonian density into
the form
\begin{eqnarray}
\mH=N^\tau \mH_\tau+N^\sigma \mH_\sigma \ ,
\nonumber \\
\end{eqnarray}
where
\begin{eqnarray}\label{mHtaumin}
& &\mH_\tau=\pi_\mu \hG^{\mu\nu}\pi_\nu+T\pi_\mu \hG^{\mu y}
(\tlambda^+\bY^++\tlambda^-\bY^-)+\nonumber \\
& &+\frac{T^2}{4}(\tlambda^+\bY^++\tlambda^-\bY^-)
\hG^{yy}(\tlambda^+\bY^++\tlambda^-\bY^-)-
T^2\tlambda^+\partial_\sigma y \bY^++T^2\tlambda^-
\partial_\sigma y\bY^-+T^2\tlambda^+\tlambda^-+\nonumber \\
& &+T^2\partial_\sigma x^\mu \hG_{\mu\nu}\partial_\sigma x^\nu \ , \quad \mH_\sigma=
p_\mu\partial_\sigma x^\mu \ .
\nonumber \\
\end{eqnarray}
Now we are ready to proceed to T-duality transformation that,
according to the general discussion presented in previous section,
is given by following transformations
\begin{equation}
p_y=-T\partial_\sigma \ty \ , \quad p_{\ty}=-T\partial_\sigma y
\end{equation}
so that
\begin{equation}
\pi_y=-T\partial_\sigma \ty+T\hat{B}_{yi}\partial_\sigma x^i=-T\bV \ , \quad
\pi_i=k_i-\hat{B}_{iy}p_{\ty} \ .
\end{equation}
Inserting this result into (\ref{mHtaumin}) we obtain T-dual
Hamiltonian constraint in the form
\begin{eqnarray}
& &\mH_\tau^T=\mH_\tau(p_y=-T\partial_\sigma \ty,
\partial_\sigma y=-T^{-1}p_{\ty})=
\nonumber \\
& &=(k_i-\hat{B}_{iy}p_{\ty})\hG^{ij}(k_j-\hat{B}_{jy}p_{\ty})-2T
(k_i-\hat{B}_{iy}p_{\ty})\hG^{iy}\bV+T^2 \bV^2\hG^{yy}+ \nonumber \\
& &+T(k_i-\hat{B}_{iy}p_{\ty})\hG^{iy}(\tlambda^+\bY^++\tlambda^-\bY^-)-
T^2\bV\hG^{yy}(\tlambda^+\bY^++\tlambda^-\bY^-)+\nonumber \\
& &+\frac{T^2}{4}(\tlambda^+\bY^++\tlambda^-\bY^-)
\hG^{yy}(\tlambda^+\bY^++\tlambda^-\bY^-)+
T\tlambda^+p_{\ty} \bY^+-T\tlambda^-
p_{\ty} \bY^-+T^2\tlambda^+\tlambda^-+\nonumber \\
& &+T^2\partial_\sigma x^i \hG_{ij}\partial_\sigma x^j
-2T\partial_\sigma x^i\hG_{iy}p_{\ty}+
p_{\ty}\hG_{yy}p_{\ty}  \ .
\nonumber \\
\end{eqnarray}
As the next step we derive corresponding Lagrangian density. With the help of  Hamiltonian
$H^T=\int d\sigma (N^\tau \mH_\tau^T+\mH_\sigma^T)$ we obtain following equations
of motion
\begin{eqnarray}
& &X^i=
2N^\tau \hG^{ij}(k_j-\hat{B}_{jy}p_{\ty})-2N^\tau T
\hG^{iy}\bV+TN^\tau \hG^{iy}(\tlambda^+\bY^++\tlambda^-\bY^-) \ , \nonumber \\
& &Y=-\hat{B}_{iy}X^i+N^\tau T (\tlambda^+\bY^+-
\tlambda^-\bY^-)-2TN^\tau \partial_\sigma x^i\hG_{iy}+2N^\tau \hG_{yy}p_{\ty}
 \ ,
\nonumber \\
\end{eqnarray}
where
\begin{equation}
 X^i=\partial_\tau x^i-N^\sigma \partial_\sigma x^i \ , \quad
 Y=\partial_\tau \ty-N^\sigma \partial_\sigma \ty \ .
 \end{equation}
Now we see that we can express $p_{\ty}$ from the last expression as
\begin{equation}
p_{\ty}=\frac{1}{2N^\tau\hG_{yy}}
(Y+\hat{B}_{iy}X^i-N^\tau T (\tlambda^+\bY^+-
\tlambda^-\bY^-)+2TN^\tau \partial_\sigma x^i\hG_{iy}) \ .
\end{equation}
To proceed further we have  to introduce metric inverse to $\hG^{ij}$. It is easy to see that it has the form
\begin{equation}
\tilde{G}_{ij}=\hG_{ij}-\frac{1}{\hG_{yy}}\hG_{iy}\hG_{yj}
\end{equation}
that obeys
\begin{equation}
\tG_{ij}\hG^{jk}
=\delta_i^k \  , \quad \tG_{ij}\hG^{jy}=-\frac{\hG_{iy}}{\hG_{yy}} \ .
\end{equation}
Then we find
\begin{equation}
(k_i-\hat{B}_{iy}p_{\ty})=
\frac{1}{2N^\tau}\tG_{ij}(X^j+2N^\tau T \hG^{jy}\bV-T\hG^{jy}N^\tau (\tlambda^+\bY^++\tlambda^-\bY^-)) \
\end{equation}
and hence after some calculations we obtain T-dual Lagrangian density in the form
\begin{eqnarray}
& &\mL^T=p_i\partial_\tau x^i+p_{\ty}\partial_\tau \ty-N^\tau \mH^T_\tau-N^\sigma
\mH^T_\sigma=\nonumber \\
& &=\frac{1}{4N^\tau}(g'_{\tau\tau}-2N^\sigma g'_{\tau\sigma}+(N^\sigma)^2
g'_{\sigma\sigma})-N^\tau T^2 g'_{\sigma\sigma}-T B_{\mu\nu}\partial_{\tau}\tx^\mu
\partial_\sigma \tx^\nu +\nonumber \\
& &+\frac{T}{2}N^\tau\tlambda^+(\nabla_n x^i\frac{\bY^+}{\hG_{yy}}(\hG_{iy}-\hat{B}_{iy})-\nabla_n\ty
\frac{\bY^+}{\hG_{yy}}+2T\partial_\sigma \ty \frac{\bY^+}{\hG_{yy}}-2T\frac{\bY^+}{\hG_{yy}}(\hG_{iy}-\hat{B}_{iy})\partial_\sigma x^i)
\nonumber \\
& &+\frac{T}{2}N^\tau\tlambda^-(\nabla_n x^i\frac{\bY^-}{\hG_{yy}}(\hG_{iy}+
\hat{B}_{iy})+\nabla_n y\frac{\bY^-}{\hG_{yy}}+2T\partial_\sigma \ty
\frac{\bY^-}{\hG_{yy}}+2T\partial_\sigma x^i\frac{\bY^-}{\hG_{yy}}(\hG_{iy}+\hat{B}_{iy}))
-\nonumber \\
& &-N^\tau T^2\tlambda^+\tlambda^-(1+\frac{1}{\hG_{yy}}\bY^+\bY^-) \ , \nonumber \\
\nonumber \\
\end{eqnarray}
where
\begin{equation}\label{gbar}
g'_{\alpha\beta}=G'_{\mu\nu}\partial_\alpha \tx^\mu \partial_\beta \tx^\nu \ , \quad
\tx^\mu\equiv (x^i,\ty)
\end{equation}
and where
\begin{eqnarray}\label{Buschrul}
& &G'_{ij}=\hG_{ij}-\frac{1}{\hG_{yy}}\hG_{iy}\hG_{yj}+
\frac{1}{\hG_{yy}}\hat{B}_{iy}\hat{B}_{jy} \ ,  \nonumber \\
& &
G'_{i\ty}=\frac{\hat{B}_{iy}}{\hG_{yy}} \ , \quad
G'_{\ty j}=-\frac{\hat{B}_{yj}}{\hG_{yy}} \ , \quad
G'_{\ty\ty}=\frac{1}{\hG_{yy}} \nonumber \\
& &
B'_{ij}=B_{ij}-\frac{\hG_{iy}}{\hG_{yy}}\hat{B}_{yj}-\frac{\hat{B}_{iy}}{\hG_{yy}}\hG_{yj} \ ,
\quad
B'_{ i\ty}=\frac{\hG_{iy}}{\hG_{yy}} \ , \quad B'_{\ty j}=-\frac{\hG_{yj}}{\hG_{yy}} \
\nonumber \\
\end{eqnarray}
that are standard Buscher's rules \cite{Buscher:1987sk,Buscher:1987qj}. However the
crucial  point of the T-dual Lagrangian is an absence of the term proportional to
$\tlambda^+\tlambda^-$ since $\bY^+\bY^-=-\hG_{yy}$ from definition.
In fact, let us introduce more compact notation
\begin{equation}
\bA'_\mu=\left(\frac{\bY^+}{\hG_{yy}}(\hG_{iy}-\hat{B}_{iy}),-\frac{\bY^+}{\hG_{yy}}\right) \ ,
\quad
\bB'_\mu=\left(\frac{\bY^-}{\hG_{yy}}(\hG_{iy}+\hat{B}_{iy}),\frac{\bY^-}{\hG_{yy}}\right) \ ,
\end{equation}
so that the Lagrangian density can be written as
\begin{eqnarray}
& &\mL^T=
\frac{1}{4N^\tau}(g'_{\tau\tau}-2N^\sigma g'_{\tau\sigma}+(N^\sigma)^2
g'_{\sigma\sigma})-N^\tau T^2 g'_{\sigma\sigma}-T B_{\mu\nu}\partial_{\tau}\tx^\mu
\partial_\sigma \tx^\nu +\nonumber \\
& &+\frac{T}{2}N^\tau\tlambda^+(\nabla_n\tx^\mu \bA_\mu-2T\partial_\sigma \tx^\mu \bA_\mu)
+\frac{T}{2}N^\tau\tlambda^-(\nabla_n\tx^\mu \bB_\mu+2T\partial_\sigma \tx^\mu\bB_\mu) \ .
\nonumber \\
\end{eqnarray}
Now the equation of motion for $\tlambda^+,\tlambda^-$ implies
\begin{equation}\label{eqtlambda}
\nabla_n \tx^\mu \bA'_\mu=2T\partial_\sigma \tx^\mu \bA'_\mu \ , \quad
\nabla_n \tx^\mu \bB'_\mu=-2T\partial_\sigma \tx^\mu \bB'_\mu \ .
\end{equation}
We can solve these equations as follow. We multiply the first equation with $\partial_\sigma \tx^\nu\bB'_\nu$ and the second one with $\partial_\sigma \tx^\nu\bA'_\nu$ and sum so that  we obtain
\begin{eqnarray}
N^\sigma=\frac{\partial_\tau \tx^\mu
\bM_{\mu\nu}\partial_\sigma \tx^\nu }
{\partial_\sigma \tx^\mu \bM_{\mu\nu}\partial_\sigma \tx^\nu} \ , \quad
\bM_{\mu\nu}=\frac{1}{2}(\bA'_\mu\bB'_\nu+\bB'_\mu\bA'_\nu) \ ,
\nonumber \\
\end{eqnarray}
where explicitly we have
\begin{eqnarray}
\bM_{ij}=-\frac{1}{\hG_{yy}}(\hG_{iy}\hG_{jy}-\hat{B}_{iy}\hat{B}_{jy}) \ , \quad
\bM_{i\ty}=-\frac{\hG_{iy}}{\hG_{yy}} \ ,
\quad
\bM_{\ty\ty}=\frac{1}{\hG_{yy}} \ .
\nonumber \\
\end{eqnarray}
Further, if we multiply two equations given in (\ref{eqtlambda}) together we obtain
\begin{equation}
N^\tau=\frac{\sqrt{-\det \bM_{\alpha\beta}}}{2\bM_{\sigma\sigma}} \ ,
\quad
\bM_{\alpha\beta}=\partial_\alpha \tx^\mu \bM_{\mu\nu}
\partial_\beta \tx^\nu
\end{equation}
and hence final Lagrangian density has the form of non-relativistic string action
\begin{eqnarray}
\mL^T=-\frac{T}{2}\sqrt{-\det\bM}\bM^{\alpha\beta}g'_{\alpha\beta}-TB' _{\mu\nu}
\partial_\tau \tx^\mu\partial_\sigma \tx^\nu \ ,
\nonumber \\
\end{eqnarray}
where $\bM^{\alpha\beta}$ is matrix inverse to $\bM_{\alpha\beta}$ so that
$\bM^{\alpha\beta}\bM_{\beta\gamma}=\delta^\alpha_\beta$. This is very interesting
result that shows that T-dual of the string in the background with light-like isometry
is non-relativistic string in dual background. In the next section we will
study this problem for more general form of terms proportional to Lagrange multipliers.
\subsection{The case of non-zero $\bA_i,\bB_i$}
Now we return to the case when $\bA_i$ and $\bB_i$ are not equal to zero. Recall
that  extended Lagrangian density has the form
\begin{eqnarray}\label{mLextendedgen}
& &\mL=-\frac{T}{2}N\sqrt{\omega}
[-\nabla_n x^i \hG_{ij}\nabla_n x^j+\frac{1}{\omega}
\partial_\sigma x^i\partial_\sigma x^j\hG_{ij}-2\nabla_n x^i
\hG_{iy}\nabla_n y+\frac{2}{\omega}\partial_\sigma x^i\hG_{iy}\partial_\sigma y-\nonumber \\
& &-\nabla_n y \hG_{yy}\nabla_n y+\frac{1}{\omega}
\partial_\sigma y \hG_{yy}\partial_\sigma y+
\lambda^+ \bA+\lambda^-\bB+\lambda^+\lambda^-]-T\hat{B}_{\mu\nu}
\nabla_n x^\mu \partial_\sigma x^\nu \ ,
\nonumber \\
\end{eqnarray}
where
\begin{eqnarray}
\bA=\nabla_n x^i \bA_i+\nabla_n y \bY^+-\frac{1}{\sqrt{\omega}}
[\partial_\sigma x^i\bA_i+\partial_\sigma y \bY^+] \ ,
\nonumber \\
\bB=\nabla_n x^i \bB_i+\nabla_n y \bY^-+\frac{1}{\sqrt{\omega}}
[\partial_\sigma x^i\bB_i+\partial_\sigma y \bY^-] \ .
\nonumber \\
\end{eqnarray}
In order to find relation with the original Lagrangian in the background with
light-like isometry let us solve the equations of motions
 $\lambda^+$ and
$\lambda^-$ and insert the results into (\ref{mLextendedgen}). Explicitly
we find that the contribution is equal to
\begin{eqnarray}
& &-\bA \bB=
-[\nabla_n x^i\nabla_n x^j\bA_i\bB_j+\nabla_n x^i\nabla_n y
(\bA_i\bY^-+\bY^+\bB_i)+\nonumber \\
& &+\frac{2}{
    \sqrt{\omega}}\nabla_n x^i\partial_\sigma x^j(\bA_i\bB_j-\bA_i\bB_j)
+\frac{1}{\sqrt{\omega}}\nabla_n x^i \partial_\sigma y(\bY^-\bA_i-
\bY^+\bB_i)+
\nonumber \\
& &+\nabla_n y\nabla_n y \bY^+\bY^-+\frac{1}{\sqrt{\omega}}\nabla_n \bY\partial_\sigma x^i
(\bY^+\bB_i-\bY^-\bA_i)
-\nonumber \\
& &-\frac{1}{\omega}(\partial_\sigma x^i\partial_\sigma x^j \bA_i\bB_j+
\partial_\sigma x^i\partial_\sigma y(\bA_i\bY^-+\bB_i\bY^+)+
\partial_\sigma y\partial_\sigma y \bY^+\bY^-)] \ .
\nonumber \\
\end{eqnarray}
First of all we demand that
\begin{equation}
\hG_{yy}+\bY^+\bY^-=0 \
\end{equation}
that can be again  solved as $\bY^+=\sqrt{\hG_{yy}} \ , \bY^-=-\sqrt{\hG_{yy}}$. Further, since
the mixed terms $\nabla_n y\partial_\sigma x^j$ should
be zero we have to demand that (since $\bY^+=-\bY^-$)
\begin{equation}
\bB_i=-\bA_i \ .
\end{equation}
Then we obtain
\begin{eqnarray}
& &-\bA \bB=
-[\nabla_n x^i\nabla_n x^j\bA_i\bB_j+2\nabla_n x^i\nabla_n y
\bA_i\bY^-
\nonumber \\
& &-\nabla_n y\nabla_n y \bY^+\bY^+
-\frac{1}{\omega}(\partial_\sigma x^i\partial_\sigma x^j \bA_i\bB_j+
2\partial_\sigma x^i\partial_\sigma y\bA_i\bY^-+
\partial_\sigma y\partial_\sigma y \bY^+\bY^-)]
\nonumber \\
\end{eqnarray}
that implies following relation between components of original and hatted metrics and
NSNS two forms
\begin{equation}\label{hGrel}
\hG_{ij}=G_{ij}+\bA_i\bA_j \ , \quad \hG_{iy}-\bA_i\sqrt{\hG_{yy}}=G_{iy} \ , \quad
\hB_{\mu\nu}=B_{\mu\nu} \ .
\end{equation}
For further purposes we introduce notation
\begin{eqnarray}
& &\bA=\nabla_n x^\mu \bA_\mu-\frac{1}{\sqrt{\omega}}\partial_\sigma x^\mu \bA_\mu \ ,
\quad  \bA_\mu=(\bA_i,\bY^+) \ , \nonumber \\
& &\bB=\nabla_n x^\mu\bB_\mu+\frac{1}{\sqrt{\omega}}\partial_\sigma x^\mu \bB_\mu \ , \quad
\bB_\mu=(\bB_i,\bY^-) \ , \quad  x^\mu=(x^i,y) \ .
\nonumber \\
\end{eqnarray}
Now we are ready to proceed to the Hamiltonian formulation of the Lagrangian
density (\ref{mLextendedgen}) when we have following conjugate momenta
\begin{eqnarray}
& &p_\mu=T\sqrt{\omega}\hG_{\mu\nu}\nabla_n x^\nu-\frac{T}{2}\sqrt{\omega}
(\lambda^+\bA_\mu+\lambda^-\bB_\mu) -T\hat{B}_{\mu\nu}\partial_\sigma x^\nu \ ,
\nonumber \\
& & p_{N}\approx 0 \ , \quad  p_{N^\sigma}\approx  0 \ , \quad p_{\lambda^+}\approx 0 \ ,
\quad
p_{\lambda^-}\approx 0 \ . \nonumber \\
\end{eqnarray}
Then performing the same analysis as in previous section we obtain Hamiltonian density in the form
\begin{equation}
\mH=N^\tau \mH_\tau+N^\sigma \mH_\sigma \ ,
\end{equation}
where
\begin{eqnarray}
\mH_\tau=\pi_\mu \hG^{\mu\nu}\pi_\nu+T^2\partial_\sigma x^\mu
\hG_{\mu\nu}\partial_\sigma x^\nu
+T\pi_\mu \hG^{\mu\nu}
(\tlambda^+\bA_\nu+\tlambda^-\bB_\nu)
+\nonumber \\
+T^2\partial_\sigma x^\mu (\tlambda^-\bB_\mu-\tlambda^+\bA_\mu)
+\frac{T^2}{4}
(\tlambda^+\bA_\mu+\tlambda^-\bA_\mu)\hG^{\mu\nu}
(\tlambda^+\bA_\nu+\tlambda^-\bB_\nu)
+T^2\tlambda^+\tlambda^- \ .
\nonumber \\
\end{eqnarray}
Performing the same T-duality transformation as in previous section we obtain T-dual
Hamiltonian
\begin{eqnarray}
& &\mH^T=N^\tau \mH_\tau^T+N^\sigma \mH_\sigma^T  \ , \quad
 \mH^T_\sigma
=p_i\partial_\sigma x^i+p_{\ty}\partial_\sigma \ty \ , \nonumber \\
& &\mH_\tau^T=(k_i-\hB_{iy}p_{\ty})\hG^{ij}
(k_j-\hB_{jy}p_{\ty})-2T\bV\hG^{yi}(k_i-\hB_{iy}p_{\ty})+
T^2\bV^2\hG^{yy}+\nonumber \\
& &+T(k_i-\hB_{iy}p_{\ty})\hG^{i\nu}(\tlambda^+\bA_\nu+\tlambda^-
\bB_\nu)-T^2\bV\hG^{y\nu}
(\tlambda^+\bA_\nu+\tlambda^-\bB_\nu)+\nonumber \\
& &+T^2\partial_\sigma x^i(\tlambda^-\bB_i-\tlambda^+\bA_i)
-Tp_{\ty}(\tlambda^-\bB_y-\tlambda^+\bA_y)+
\nonumber \\
& &+T^2\partial_\sigma x^i\hG_{ij}\partial_\sigma x^j-2Tp_{\ty}
\hG_{yi}\partial_\sigma x^i+
p_{\ty}^2 \hG_{yy} +
\nonumber \\
& &+\frac{T^2}{4}(\tlambda^+\bA_\mu+\tlambda^-\bB_\mu)
\hG^{\mu\nu}(\tlambda^+\bA_\nu+\tlambda^-\bB_\nu)+T^2
\tlambda^+\tlambda^- \ . \nonumber \\
\end{eqnarray}
As the next step we again determine corresponding Lagrangian density
using equations of motion for $x^i,\ty$. Since the analysis is completely
the same as in previous section we write the final result
\begin{eqnarray}
\mL^T
=\frac{1}{4N^\tau}(g'_{\tau\tau}-2N^\sigma g'_{\tau\sigma}+(N^\sigma)^2
g'_{\sigma\sigma})-N^\tau T^2 g'_{\sigma\sigma}-T B_{\mu\nu}\partial_{\tau}\tx^\mu
\partial_\sigma \tx^\nu +\nonumber \\
+\frac{T}{2}N\tlambda^+(\nabla_n\tx^\mu \bA'_\mu-2T\partial_\sigma \tx^\mu \bA'_\mu)
+\frac{T}{2}N\tlambda^-(\nabla_n\tx^\mu \bB'_\mu+2T\partial_\sigma \tx^\mu\bB'_\mu) \ ,
\nonumber \\
\end{eqnarray}
where
\begin{eqnarray}
& &\bA_\mu'= \left((\frac{\hG_{iy}}{\hG_{yy}}\bA_y-\frac{\hB_{iy}}{\hG_{yy}}\bA_y-\bA_i),-\frac{\bA_y}
{\hG_{yy}}\right) \ , \nonumber \\
& &\bB_\mu'=\left((\frac{\hG_{iy}}{\hG_{yy}}\bB_y
+\frac{\hB_{iy}}{\hG_{yy}}\bB_y
-\bB_i),\frac{\bB_y}{\hG_{yy}}\right) \ , \nonumber \\
\end{eqnarray}
and where $g'_{\mu\nu}$ was given in (\ref{gbar}) and the background
fields in (\ref{Buschrul}).  As the final step we solve the equations of motion
for $\tlambda^+,\tlambda^-$ and we obtain final form of the Lagrangian density
for T-dual theory
%
\begin{eqnarray}
\mL^T=-\frac{T}{2}\sqrt{-\det\bM}\bM^{\alpha\beta}g'_{\alpha\beta}-TB' _{\mu\nu}
\partial_\tau \tx^\mu\partial_\sigma \tx^\nu \ , \quad  \bM_{\alpha\beta}=\partial_\alpha \tx^\mu \bM_{\mu\nu}\partial_\beta \tx^\nu \ ,
\nonumber \\
\end{eqnarray}
where
\begin{eqnarray}
& &\bM_{\ty\ty}=\frac{1}{\hG_{yy}} \ , \quad \bM_{i\ty}=-\frac{\hG_{iy}}{\hG_{yy}} \ ,
\nonumber \\
& &\bM_{ij}=-\frac{1}{\hG_{yy}}(\hG_{iy}\hG_{jy}-B_{iy}B_{jy})-\bA_i\bA_j+\frac{\hG_{iy}}{\hG_{yy}}\bA_j+\frac{\hG_{jy}}{\hG_{yy}}\bA_i \ .
\nonumber \\
\end{eqnarray}
We see that the presence of the more general form of $\bA_\mu,\bB_\mu$ only affects
the form of the matrix $\bM_{\mu\nu}$. On the other hand since $\bA_i$ are determined
by the equation (\ref{hGrel}) and the choice of the  metric components $\hG_{iy}$ we
see that it is more convenient to consider minimal case when $\bA_i=0$.

Let us outline our results. We analyzed T-duality of relativistic string along light-like
isometry and we argued that T-dual theory has the form of non-relativistic string action
on T-dual background.



\begin{thebibliography}{20}



\bibitem{Buscher:1987sk}
T.~H.~Buscher,
\emph{``A Symmetry of the String Background Field Equations,''}
    Phys.\ Lett.\ B {\bf 194} (1987) 59.
    doi:10.1016/0370-2693(87)90769-6

\bibitem{Buscher:1987qj}
T.~H.~Buscher,
\emph{``Path Integral Derivation
    of Quantum Duality in Nonlinear Sigma Models,''}
Phys.\ Lett.\ B {\bf 201} (1988) 466.
doi:10.1016/0370-2693(88)90602-8


    \bibitem{Alvarez:1994wj}
    E.~Alvarez, L.~Alvarez-Gaume and Y.~Lozano,
    \emph{``A Canonical approach to duality transformations,''}
    Phys.\ Lett.\ B {\bf 336} (1994) 183
    doi:10.1016/0370-2693(94)00982-1
    [hep-th/9406206].

    \bibitem{Alvarez:1994dn}
    E.~Alvarez, L.~Alvarez-Gaume and Y.~Lozano,
    \emph{``An Introduction to T duality in string theory,''}
    Nucl.\ Phys.\ Proc.\ Suppl.\  {\bf 41} (1995) 1
    doi:10.1016/0920-5632(95)00429-D
    [hep-th/9410237].

\bibitem{Bergshoeff:2018yvt}
E.~Bergshoeff, J.~Gomis and Z.~Yan,
\emph{``Nonrelativistic String Theory and T-Duality,''}
JHEP {\bf 1811} (2018) 133
[arXiv:1806.06071 [hep-th]].

\bibitem{Kluson:2018vfd}
J.~Kluso\v{n},
\emph{``Note About T-duality of Non-Relativistic String,''}
arXiv:1811.12658 [hep-th].







\bibitem{Gomis:2000bd}
J.~Gomis and H.~Ooguri,
\emph{``Nonrelativistic closed string theory,''}
J.\ Math.\ Phys.\  {\bf 42} (2001) 3127
doi:10.1063/1.1372697
[hep-th/0009181].

\bibitem{Danielsson:2000gi}
U.~H.~Danielsson, A.~Guijosa and M.~Kruczenski,
\emph{``IIA/B, wound and wrapped,''}
JHEP {\bf 0010} (2000) 020
doi:10.1088/1126-6708/2000/10/020
[hep-th/0009182].






\bibitem{Gomis:2005pg}
J.~Gomis, J.~Gomis and K.~Kamimura,
\emph{``Non-relativistic superstrings: A New soluble sector of AdS(5) x S**5,''}
JHEP {\bf 0512} (2005) 024
doi:10.1088/1126-6708/2005/12/024
[hep-th/0507036].





\bibitem{Gomis:2019zyu}
J.~Gomis, J.~Oh and Z.~Yan,
\emph{``Nonrelativistic String Theory in Background Fields,''}
arXiv:1905.07315 [hep-th].


\bibitem{Andringa:2012uz}
R.~Andringa, E.~Bergshoeff, J.~Gomis and M.~de Roo,
\emph{``'Stringy' Newton-Cartan Gravity,''}
Class.\ Quant.\ Grav.\  {\bf 29} (2012) 235020
doi:10.1088/0264-9381/29/23/235020
[arXiv:1206.5176 [hep-th]].






\bibitem{Harmark:2017rpg}
T.~Harmark, J.~Hartong and N.~A.~Obers,
\emph{``Nonrelativistic strings and limits of the AdS/CFT correspondence,''}
Phys.\ Rev.\ D {\bf 96} (2017) no.8,  086019
doi:10.1103/PhysRevD.96.086019
[arXiv:1705.03535 [hep-th]].



\bibitem{Kluson:2018egd}
J.~Kluso\v{n},
\emph{``Remark About Non-Relativistic
    String in Newton-Cartan Background and Null Reduction,''}
JHEP {\bf 1805} (2018) 041
doi:10.1007/JHEP05(2018)041
[arXiv:1803.07336 [hep-th]].


\bibitem{Kluson:2018grx}
J.~Kluso\v{n},
\emph{``Nonrelativistic String
    Theory Sigma Model and Its Canonical Formulation,''}
Eur.\ Phys.\ J.\ C {\bf 79} (2019) no.2,  108
doi:10.1140/epjc/s10052-019-6623-9
[arXiv:1809.10411 [hep-th]].



\bibitem{Harmark:2018cdl}
T.~Harmark, J.~Hartong, L.~Menculini, N.~A.~Obers and Z.~Yan,
\emph{``Strings with Non-Relativistic
    Conformal Symmetry and Limits of the AdS/CFT Correspondence,''}
JHEP {\bf 1811} (2018) 190
doi:10.1007/JHEP11(2018)190
[arXiv:1810.05560 [hep-th]].

\bibitem{Kluson:2019ifd}
J.~Kluso\v{n},
\emph{``$(m,n)$-String and D1-Brane in Stringy Newton-Cartan Background,''}
JHEP {\bf 1904} (2019) 163
doi:10.1007/JHEP04(2019)163
[arXiv:1901.11292 [hep-th]].
















\bibitem{Kluson:2018uss}
J.~Kluso\v{n},
\emph{``Hamiltonian for a string in a Newton-Cartan background,''}
Phys.\ Rev.\ D {\bf 98} (2018) no.8,  086010
doi:10.1103/PhysRevD.98.086010
[arXiv:1801.10376 [hep-th]].







\end{thebibliography}
\end{document}